\documentclass{moriond}

\def\Journal#1#2#3#4{{#1} {\bf #2}, #3 (#4)}

\def\PRL{\em Phys. Rev. Lett.}
\def\PRD{{\em Phys. Rev.} D}

\usepackage{lineno} 
\usepackage{amsmath} 
\usepackage{amssymb} 

\def\be{\begin{equation}}
\def\ee{\end{equation}}
\def\bea{\begin{eqnarray}}
\def\eea{\end{eqnarray}}

\newcommand\blfootnote[1]{
  \begingroup
  \renewcommand\thefootnote{}\footnote{#1}%
  \addtocounter{footnote}{-1}%
  \endgroup
}

\def\ifb{\ifmmode {\mathrm{\ fb}^{-1}}\else
                   \ensuremath{\mathrm{\ fb}^{-1}}\fi}
\def\fb{\ifmmode {\mathrm{\ fb}}\else
  \ensuremath{\mathrm{\ fb}}\fi}
\def\pb{\ifmmode {\mathrm{\ pb}}\else
                   \ensuremath{\mathrm{\ pb}}\fi}

\def\TeV{\ifmmode {\mathrm{\ Te\kern -0.1em V}}\else
                   \textrm{Te\kern -0.1em V}\fi }%
\def\GeV{\ifmmode {\mathrm{\ Ge\kern -0.1em V}}\else
  \textrm{Ge\kern -0.1em V}\fi }%
\newcommand{\pT}{\ensuremath{p_{\text{T}}}}
\newcommand{\etmiss}{\ensuremath{E_{\text{T}}^{\text{miss}}}}

\def\higgs{\ensuremath{H}}
\def\hihi{\ensuremath{HH}}
\def\mh{\ensuremath{m_{\higgs}}}
\def\antibar#1{\ensuremath{#1\bar{#1}}}%

\def\ttbar{\antibar{t}}%

\def\bbbar{\antibar{b}}%

\def\ccbar{\antibar{c}}%
\def\ggH{\ensuremath{\mathrm{ggF}}}
\def\VBF{\ensuremath{\mathrm{VBF}}}
\def\VH{\ensuremath{VH}}
\def\WH{\ensuremath{WH}}
\def\ZH{\ensuremath{ZH}}
\def\tth{\ensuremath{\ttbar\higgs}}
\def\th{\ensuremath{t\higgs}}

\newcommand{\Photo}{}

\begin{document}
\vspace*{4cm}
\title{HIGGS HIGHLIGHTS AT ATLAS}

\author{L. MIJOVI\'C, on behalf of the ATLAS collaboration}

\address{SUPA - School of Physics and Astronomy\\
  University of Edinburgh, United Kingdom}

\maketitle\abstracts{As the Higgs boson turns 10 this Summer, and the 
  Large Hadron Collider experiments eagerly await the next round of data-taking,
  the ATLAS collaboration has recently reached important new insights into the Higgs
  mechanism. This report is a snapshot of Higgs highlights for the MoriondQCD
  conference, obtained with 139\,\ifb{} of Run2 proton--proton collision data taken at centre-of-mass energy of 13\,\TeV.}

\section{Higgs Boson Production and Decay}

At the Large Hadron Collider (LHC) the Higgs boson is produced
through gluon-gluon fusion (\ggH) 87\% of the time, followed by vector boson
fusion (\VBF, 7\%) associated weak gauge boson production (\VH, 4\%), associated
top-antitop production (\tth, 1\%), and other production modes with small
cross-sections~\cite{yr4}. The mass of the Higgs boson is measured to be
\mh=125.09~\GeV~\cite{COMBmh}, at which the Standard Model (SM) Higgs boson decays to
\bbbar{} pairs most of the time, with a branching ratio (BR) of $\sim$58.1\%.
However, in several measurements the highest sensitivity is achieved in final states
detected with higher purity and resolution, in particular
\higgs$\rightarrow\tau\tau$ decay with BR$\sim$6.26\%,
\higgs$\rightarrow\gamma\gamma$ decay with BR$\sim$0.227\% and
\higgs$\rightarrow ZZ^* \rightarrow 4\ell$ ($\ell=e,\,\mu$) decay with a branching
ratio of only $\sim$0.0125\%~\cite{yr4}.\blfootnote{Copyright 2022 CERN for the benefit of the ATLAS Collaboration.
  CC-BY-4.0 license.} 

\section{Higgs Boson Cross-Section and Coupling Strength Measurements}
\label{sec:xc}

\begin{figure}[!ht]
  \vspace{-10pt}
  \begin{minipage}[c]{0.5\textwidth}
\centerline{\includegraphics[width=0.93\textwidth]{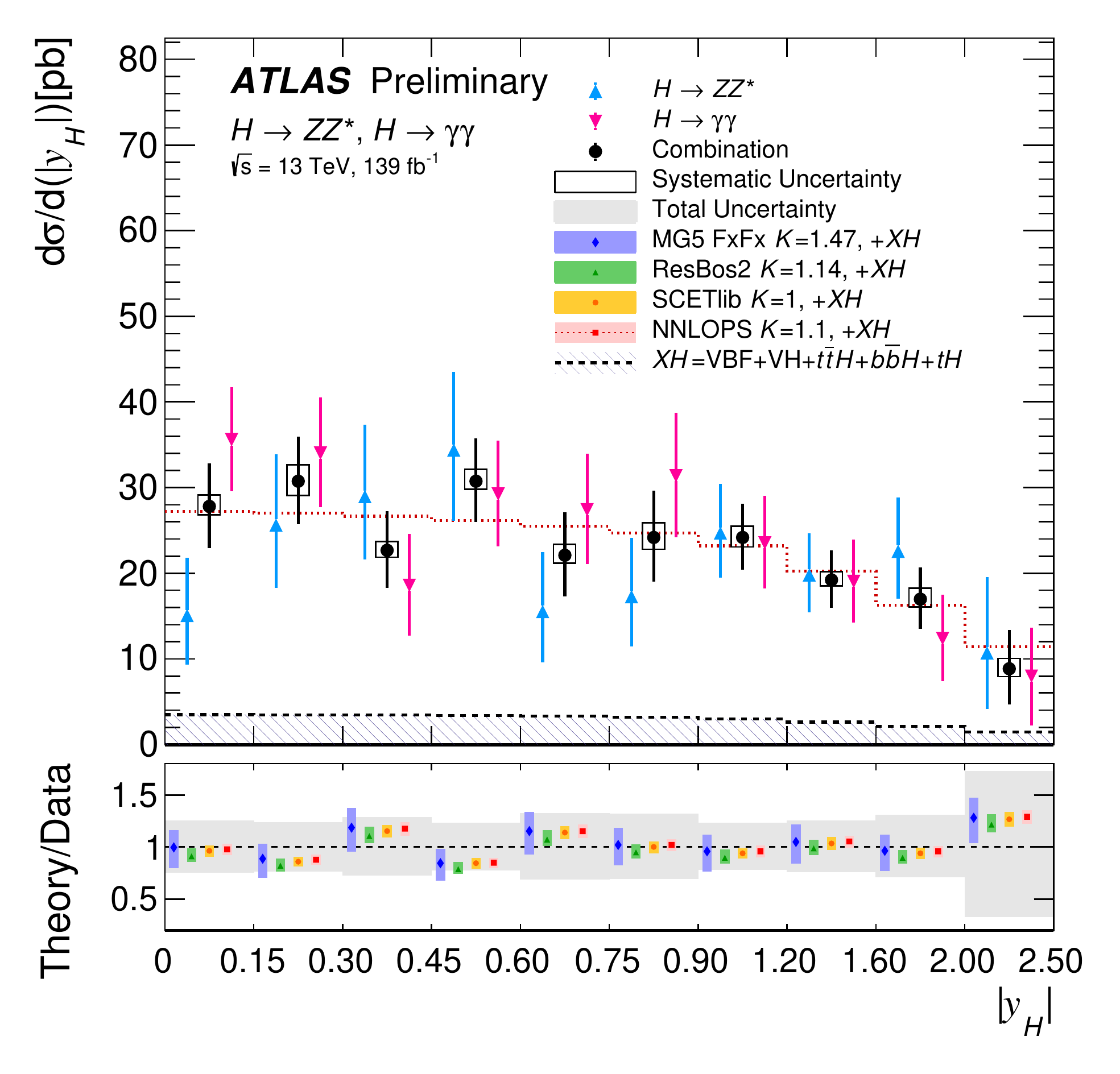}}
\end{minipage}
  \begin{minipage}[c]{0.5\textwidth}
\centerline{\includegraphics[width=0.93\textwidth]{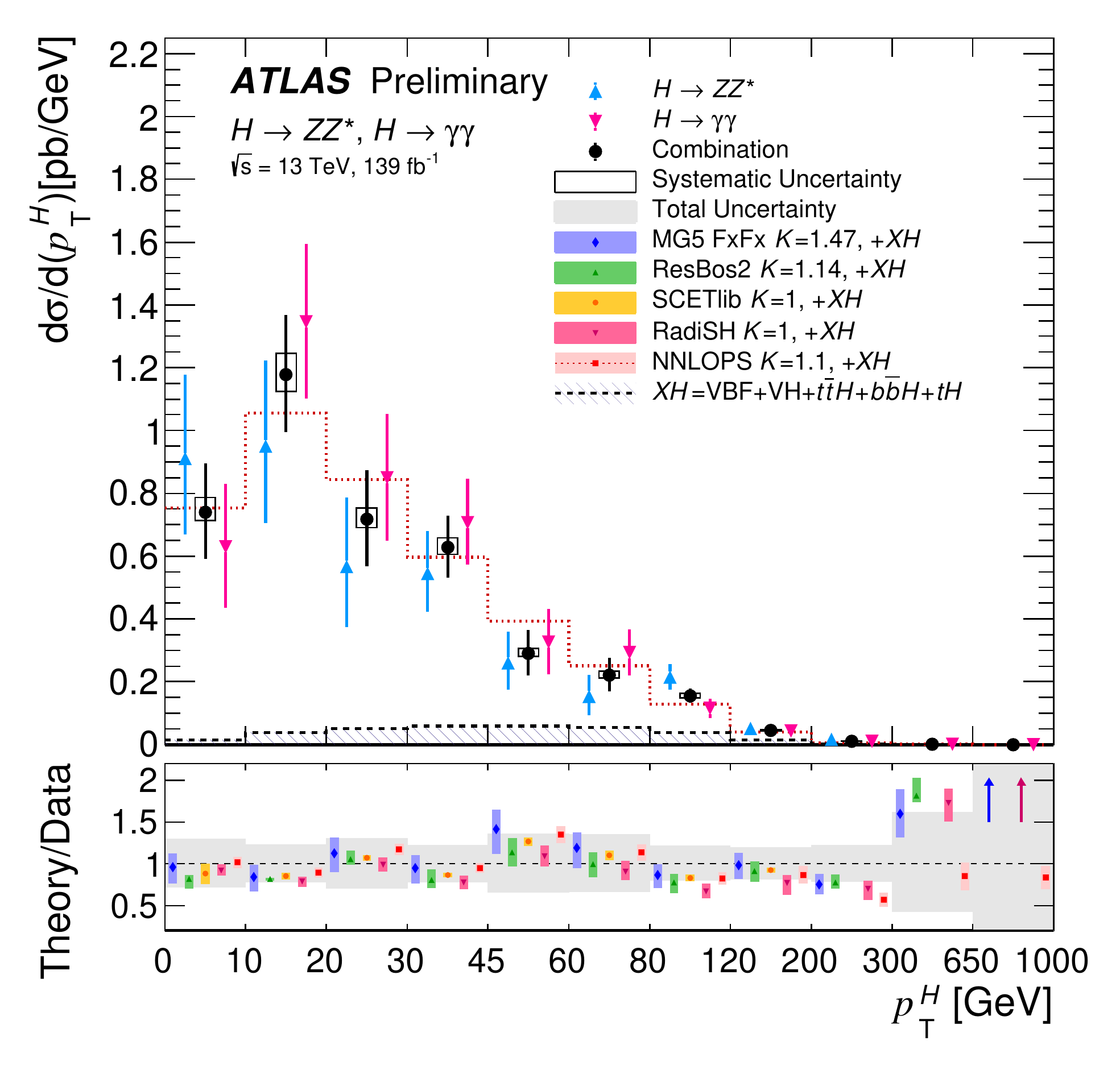}}
\end{minipage}
\vspace{-10pt}
\caption[]{Differential cross-section measurements as a function of
  $|y_{\higgs}|$ (left) and $\pT^{\higgs}$ (right) compared to the theory
  predictions for the \ggH{} production~\cite{ATLASx}.} 
\label{fig:x}
\end{figure}

With the Run2 data the uncertainty on the Higgs boson production
cross-section and coupling measurements has reached sub-10\% and sub-5\%
respectively. In a combination of \higgs$\rightarrow\gamma\gamma$ and \higgs$\rightarrow ZZ^*
\rightarrow 4\ell$ decays ATLAS\,\cite{ATLAS} has measured the inclusive cross-section 
with 7\% uncertainty: 
\begin{equation}
  \sigma(pp{\rightarrow}H) = 55.5^{+4.0}_{-3.8}\,\pb = 55.5 \pm 3.2\,\mathrm{(stat.)}~^{+2.4}_{-2.2}\,\mathrm{(sys.)}\,\pb,
\end{equation}
where $\mathrm{(stat.)}$ and $\mathrm{(sys.)}$ denote the statistical and systematic
uncertainty respectively~\cite{ATLASx}.
The measured value is compatible with the SM prediction of $55.6 \pm 2.5\,\pb$~\cite{yr4},
and the statistical uncertainty of
6\% dominates over the 4\% systematic uncertainty.   
The cross-section is also measured differentially; 
Figure~\ref{fig:x} shows the cross-sections 
as a function of the Higgs boson's rapidity $|y_{\higgs}|$
sensitive to parton distribution functions (PDF),
and the transverse momentum of the Higgs boson ($\pT^{\higgs}$)
sensitive to perturbative QCD calculations.
The measurements in the \higgs$\rightarrow\gamma\gamma$ and \higgs$\rightarrow ZZ^*
\rightarrow 4\ell$ decays are compatible with a p-value of 23\% for
$|y_{\higgs}|$ and 20\% for $\pT^{\higgs}$. The measurements are compared to the
theory predictions for the \ggH{} production, assuming SM predictions for the
production modes other than the \ggH{}. For $|y_{\higgs}|$ the combined
measurement is compatible with all the considered predictions with p-value $\geq 92\%$. The predictions are obtained
with PDF4LHC NNLO PDF (NNLOPS, ResBos2 and SCETlib) and NNPDF30 NLO PDF set (MG5FxFx). For $\pT^{\higgs}$ the
p-values range between 3.1\% (RadiSH) and 78\% (NNLOPS).  

\begin{figure}[!ht]
  \vspace{-10pt}
  \begin{minipage}[c]{0.55\textwidth}
\centerline{\includegraphics[width=1.00\textwidth]{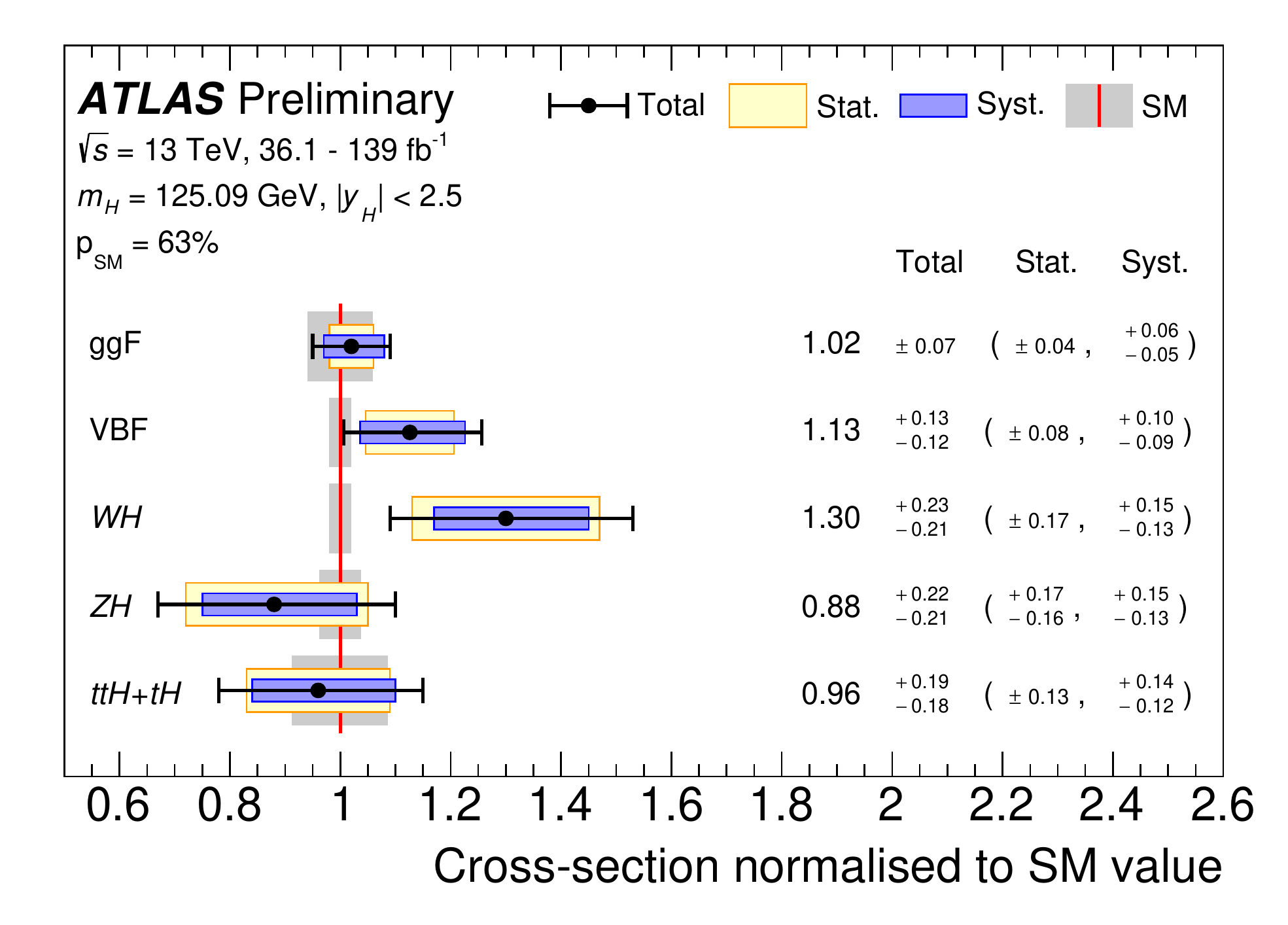}}
\end{minipage}
\hfill
\begin{minipage}[c]{0.45\textwidth}
\centerline{\includegraphics[width=1.00\textwidth]{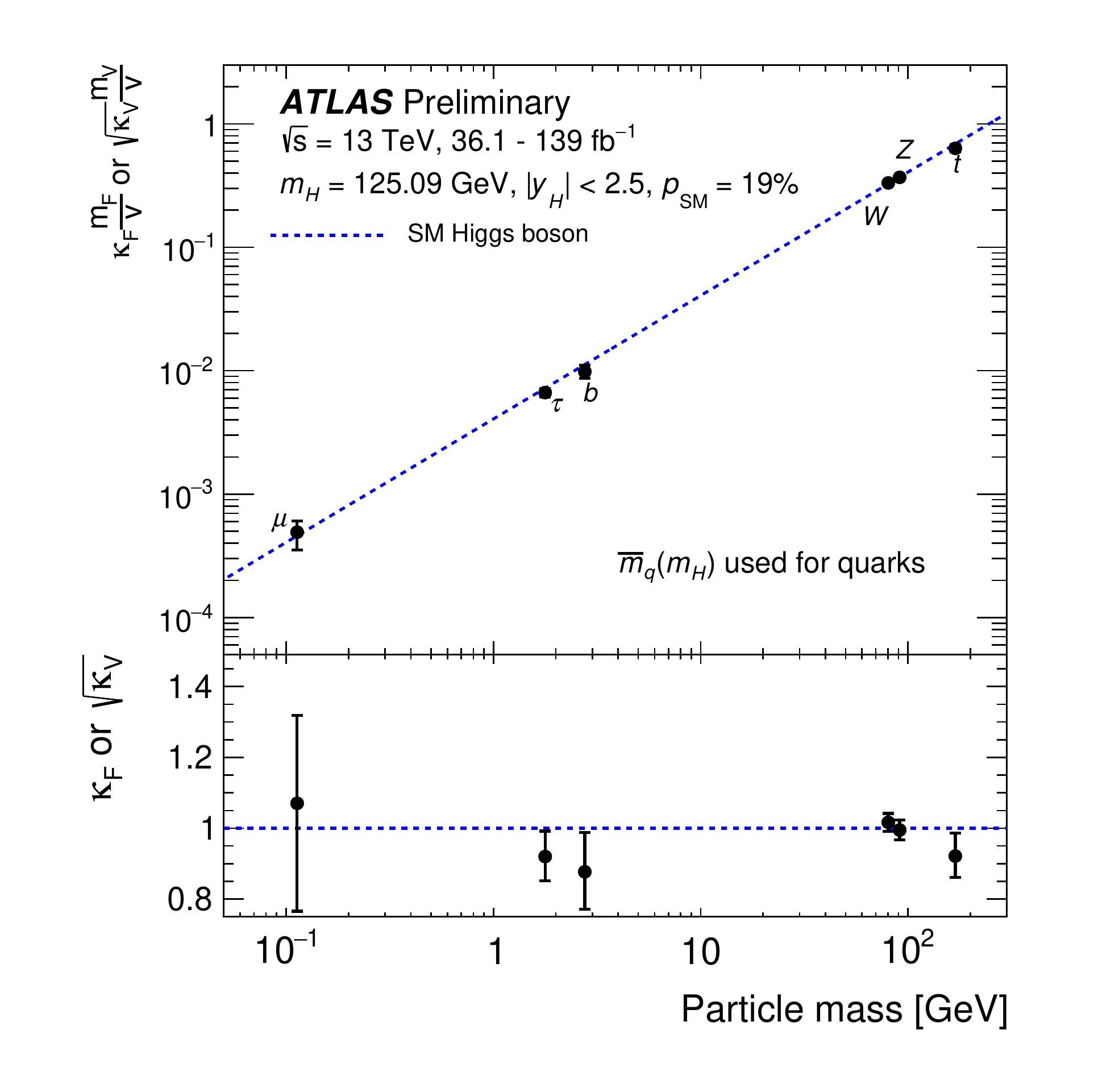}}
\end{minipage}
\vspace{-10pt}
\caption[]{Left: cross-sections for \ggH,\VBF,\WH,\ZH{} and \tth+\th{} 
  production modes. 
  Right: coupling strength modifiers for fermions F=($t,\,b,\,\tau,\,\mu$)
  and weak gauge bosons V. The dashed line shows the SM prediction~\cite{ATLASc}.}
\label{fig:comb}
\end{figure}

\noindent In the combination of
$\higgs\rightarrow\gamma\gamma,\,ZZ^*,\,WW^*,\,\tau\tau,\,\bbbar,\,\mu\mu,\,Z\gamma$ decay
modes and searches for decay into invisible final states~\cite{ATLASc}, 
the cross-sections for the main production modes have been measured with
uncertainties of $\lesssim 23\%$, and the systematic uncertainty is about
the same size as the statistical uncertainty,
as shown in Figure~\ref{fig:comb} (left).

\noindent The cross-section measurements can be interpreted in terms of coupling
strengths of the Higgs field to fermions or weak gauge bosons as shown in
Figure~\ref{fig:comb} (right). Although these coupling strengths are compatible with the SM,
there are several gaps in our experimental knowledge, due to which we
can not claim to have verified that the Higgs couplings are as predicted
by the SM. These include: (1) a model-independent extraction of the couplings.
The couplings are obtained from cross-sections, assuming no interactions between
the Higgs boson and BSM particles, and taking the width of the Higgs boson to be as
predicted by the SM; (2) measurement of the 1st and 2nd generation fermion
couplings; (3) measurement of the Higgs self-coupling. Efforts to address (1)
include Effective Field Theory interpretations and measurements of the Higgs width.  
The next two sections review the large progress ATLAS has recently made on (2)
and (3). 

\section{Search for $\boldsymbol{{\VH,{\higgs}{\rightarrow}{\ccbar}}}$
  Production}
\label{sec:vcc}

The search for ${\higgs}{\rightarrow}{\ccbar}$ decays~\cite{ATLASvcc} is
performed in \VH{} production. This is the golden production mode for measurements of
${\higgs}{\rightarrow}{\bbbar}$ decays, because the leptonic decay of the weak
gauge boson enables efficient triggering and background rejection.
In addition to challenges faced by ${\higgs}{\rightarrow}{\bbbar}$ measurements,
the ${\higgs}{\rightarrow}{\ccbar}$ analysis deals with:
\begin{itemize}
  \vspace{-8pt}
\item low branching ratio: BR$({\higgs}{\rightarrow}{\ccbar})=2.89\%$,
  about 20 times smaller than the BR(${\higgs}{\rightarrow}{\bbbar}$).
  \vspace{-8pt}
\item Charm tagging, which requires identification of charm-hadron decays with
  less distinct signature compared to $B$-hadron decays.
 \vspace{-8pt} 
\end{itemize}
Separating charm-jets from $b$-jets is particularly challenging;
this is addressed by a dedicated charm tagging algorithm,
which requires the jets to have a high 
$c$-tagging score as well as vetoing jets with high $b$-tagging scores.
This veto makes the signal events in ${\higgs}{\rightarrow}{\ccbar}$ and
${\higgs}{\rightarrow}{\bbbar}$ analyses orthogonal and enables their combination.

\begin{figure}[!ht]
    \vspace{-10pt}  
  \begin{minipage}[c]{0.50\textwidth}
\centerline{\includegraphics[width=0.95\textwidth]{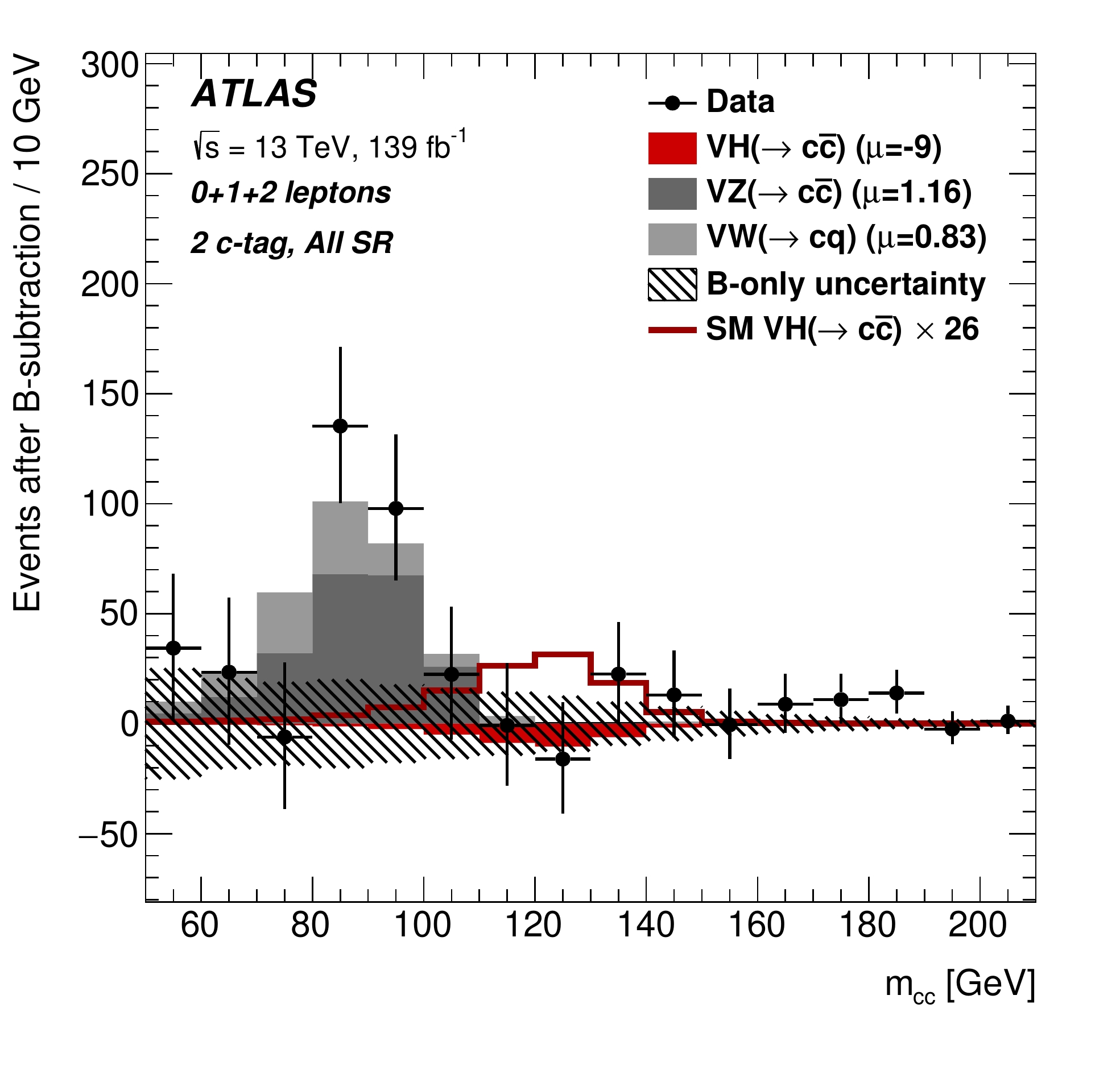}}
\end{minipage}
\hfill
\begin{minipage}[c]{0.50\textwidth}
\centerline{\includegraphics[width=0.95\textwidth]{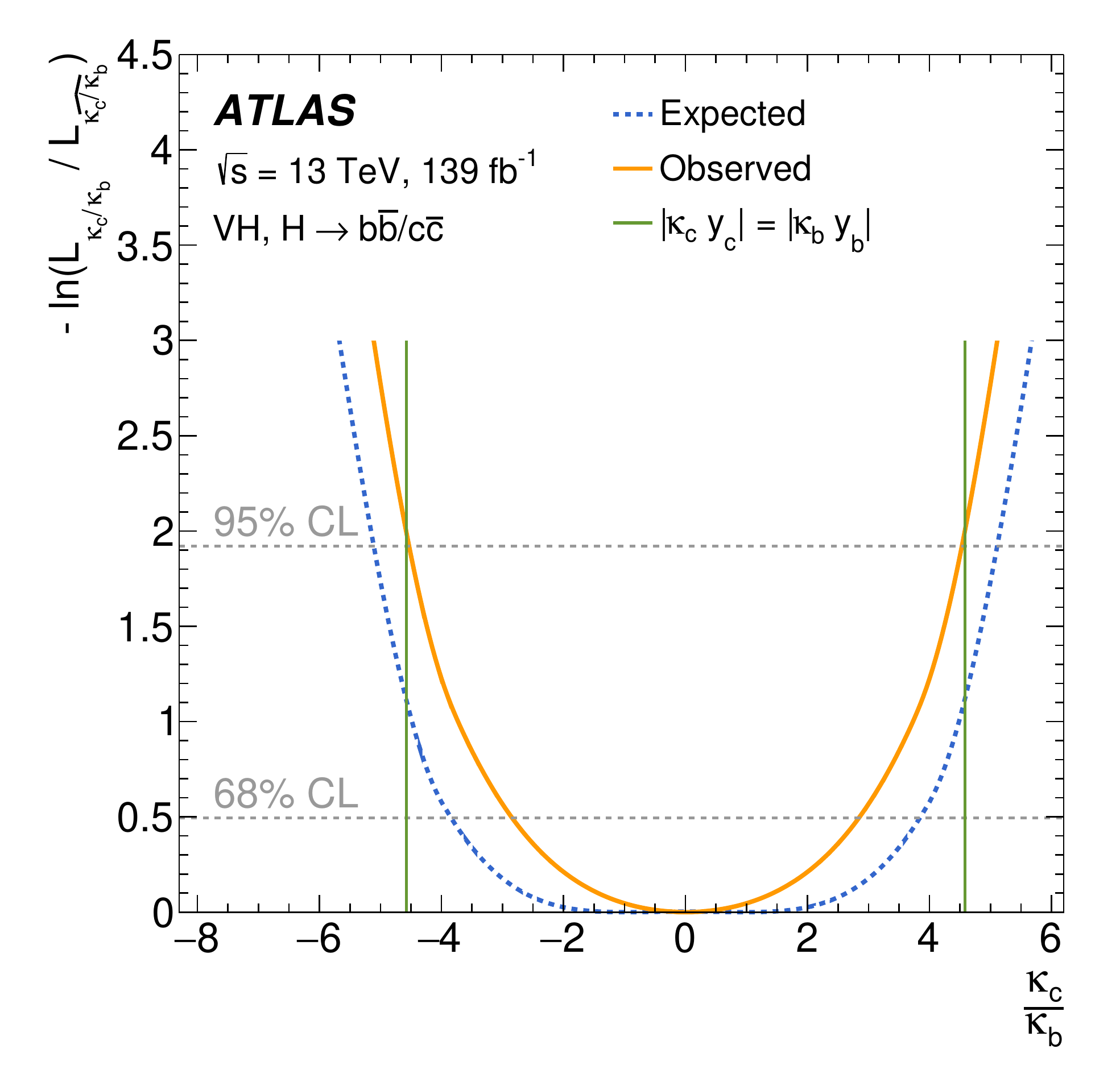}}
\end{minipage}
\vspace{-10pt}
\caption[]{Left: signal strength ($\mu$) fit to $m_{\ccbar}$ in the search for
  $\VH,{\higgs}{\rightarrow}{\ccbar}$ events. Right: constraints on coupling
  modifier ratio $|\kappa_c/\kappa_b|$ obtained in a combination of
  $\VH,{\higgs}{\rightarrow}{\ccbar}$ and $\VH,{\higgs}{\rightarrow}{\bbbar}$
  analyses~\cite{ATLASvcc}.} 
\label{fig:vcc}
\end{figure}

The $\VH,{\higgs}{\rightarrow}{\ccbar}$ analysis extracts the signal and backgrounds from 
a simultaneous fit to invariant mass of the $c$-tagged jets ($m_{\ccbar}$).  
In the fit, three signal strengths $(\mu)$, which multiply the SM
cross-sections, are extracted: $\mu_{\VH(\ccbar)}$, and the
di-boson production signal strengths $\mu_{VW(cq)}$ and $\mu_{VZ(\ccbar)}$. Figure
\ref{fig:vcc} (left) shows the best-fit distributions (solid histograms)
compared to the data. The solid line corresponds to the 95\% confidence level (CL) upper limit,
$\mu_{\VH(\ccbar)}=26$.
 
\noindent A combination of ${\higgs}{\rightarrow}{\ccbar}$ and 
${\higgs}{\rightarrow}{\bbbar}$ analyses is performed, and interpreted 
in terms of constraints on the ratio: $|\kappa_c/\kappa_b|.$
The coupling strength modifiers ($\kappa$) are defined as the ratios
of the measured Yukawa couplings ($y$) and their SM values: $\kappa_c=y_c/y_c^{\mathrm{SM}}$ and
$\kappa_b=y_b/y_b^{\mathrm{SM}}.$ Within the SM $y_b$ exceeds $y_c$ by a factor
equal to the ratios of the $b$- and $c$-quark masses: $m_b/m_c = 4.578 \pm 0.008$~\cite{vccmass}. 
Figure \ref{fig:vcc} (right) shows the observed bound: $|\kappa_c/\kappa_b|<4.5$ at 95\% CL. This is 
slightly smaller than the ratio $m_b/m_c$ shown in vertical lines, which means
that the Higgs-charm coupling is weaker than the Higgs-bottom coupling.
This important result requires no assumptions about the width of the Higgs boson, because 
the bound is extracted from the ratio of the coupling modifiers rather than
from their absolute values.  

\section{Searches for Di-Higgs Production}
\label{sec:hihi}

\begin{figure}[!ht]
  \vspace{-5pt}
  \centerline{\includegraphics[width=1.00\textwidth]{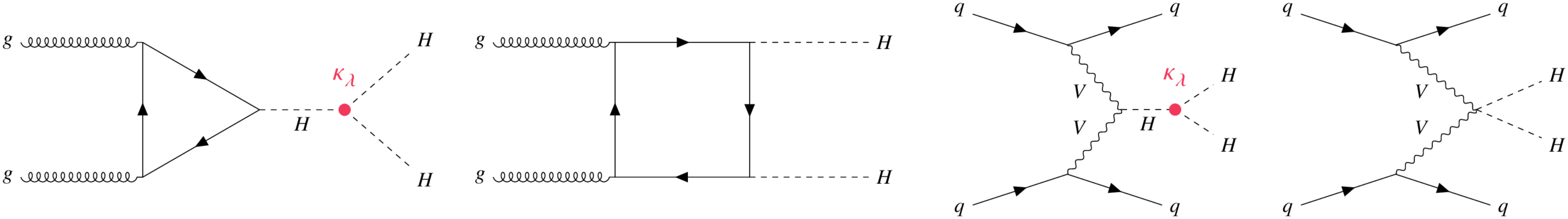}}
  \vspace{-5pt}
\caption[]{The ggF and VBF di-Higgs production diagrams.}
\label{fig:HHdia}
\end{figure}

Di-Higgs production is a probe of the Higgs potential $V$, 
which can be expanded around its minimum by an excitation ($h$) of the Higgs field
as: 
\begin{equation}
  V(h) = \frac{1}{2}\mh h^2 + \lambda_3 v h^3 + \frac{1}{4}\lambda_4 h^4.
\end{equation}
Within the SM the self-couplings are: $\lambda_3^{\mathrm{SM}} = \lambda_4^{\mathrm{SM}} = 
\mh^2/(2v^2)$. The di-Higgs searches probe the trilinear coupling $\lambda_3$, parameterised by the
coupling modifier $\kappa_\lambda= \lambda_3 / \lambda^{\mathrm{SM}}_3$.
At the LHC, the di-Higgs production proceeds through ggF $\sim$95\% of the time.
The second largest contribution is from VBF production. The corresponding
diagrams are shown in Figure~\ref{fig:HHdia}. 

\begin{figure}[!ht]
    \vspace{-10pt}
\begin{minipage}[c]{0.5\textwidth}
\centerline{\includegraphics[width=0.950\textwidth]{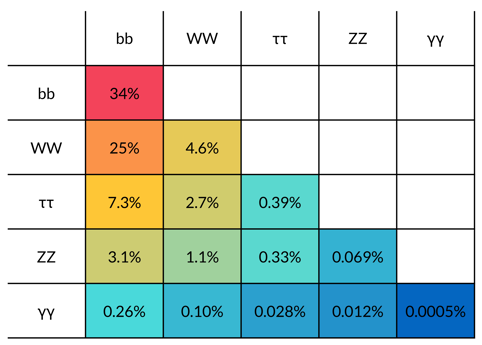}}
\end{minipage}
\hfill
\begin{minipage}[c]{0.5\textwidth}
\centerline{\includegraphics[width=0.9\textwidth]{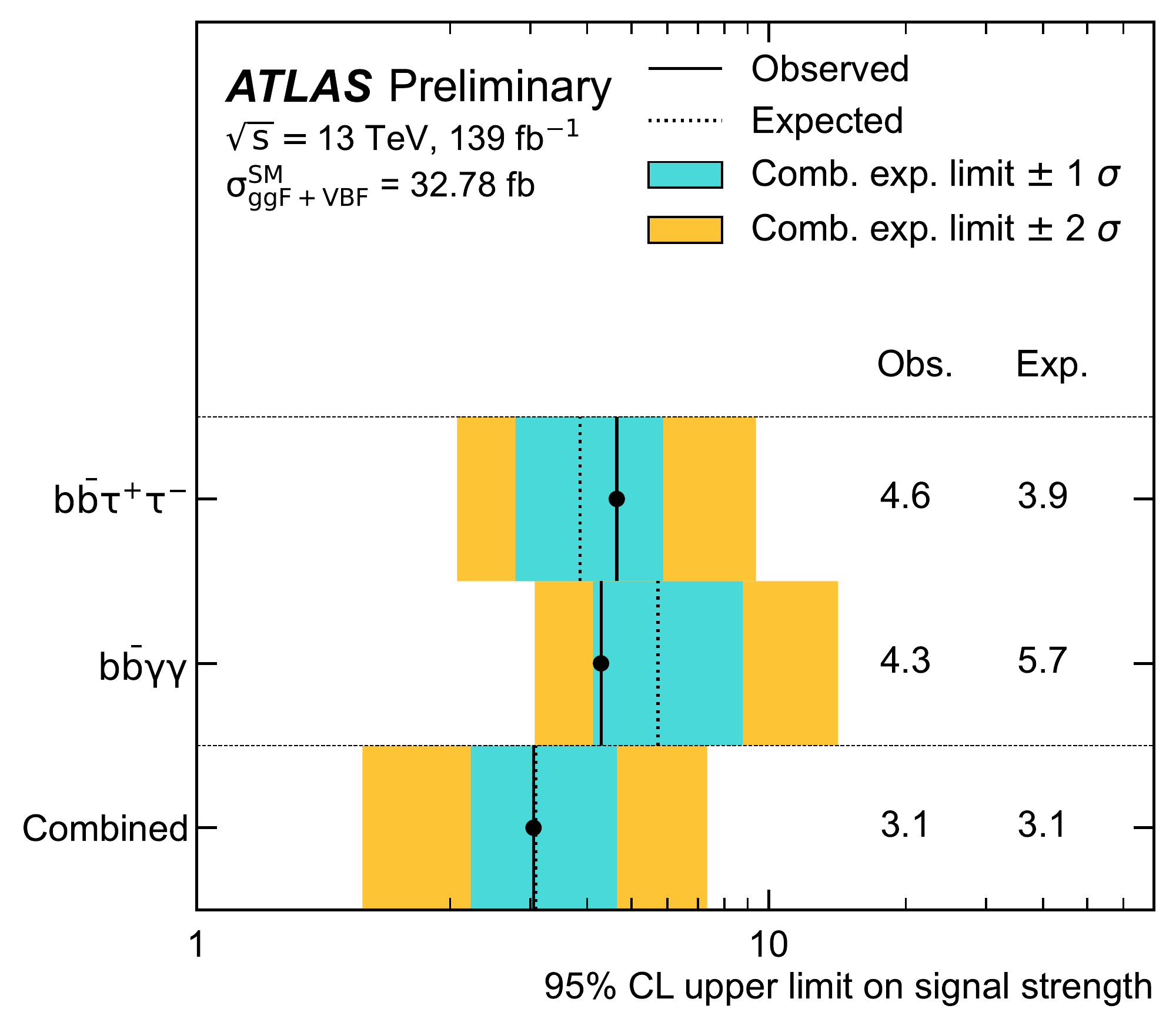}}
\end{minipage}
\vspace{-10pt}
\caption[]{Branching ratios for \hihi{} production
  (left)\
  and upper limit on the \hihi{} signal strength (right)~\cite{ATLAShh}.}
\label{fig:HHnow}
\end{figure}

\noindent The diagrams with and without the trilinear
coupling interfere destructively, leading to small cross-sections:
$\sigma_{\mathrm{ggF}}(pp\rightarrow \hihi) = 31.05 \fb$ and
$\sigma_{\mathrm{VBF}}(pp\rightarrow \hihi) = 1.73 \fb$.
Due to this, the highest sensitivity is achieved in searches in which one of the Higgs bosons
decays to \bbbar{} pair with the highest branching ratio, shown in the first
column of Figure~\ref{fig:HHnow} (left), and the other Higgs
boson decays to a distinct signature: \higgs$\rightarrow\tau\tau$ or 
\higgs$\rightarrow\gamma\gamma$. The observed and expected limits on the signal
strength $\sigma(pp{\rightarrow}HH)/\sigma(pp\rightarrow HH)^{\mathrm{SM}}$ from
these decay modes, and their combination, are shown in
Figure~\ref{fig:HHnow} (right). 

\begin{figure}[!ht]
\begin{minipage}[t]{0.515\textwidth}
\centerline{\includegraphics[width=0.95\textwidth]{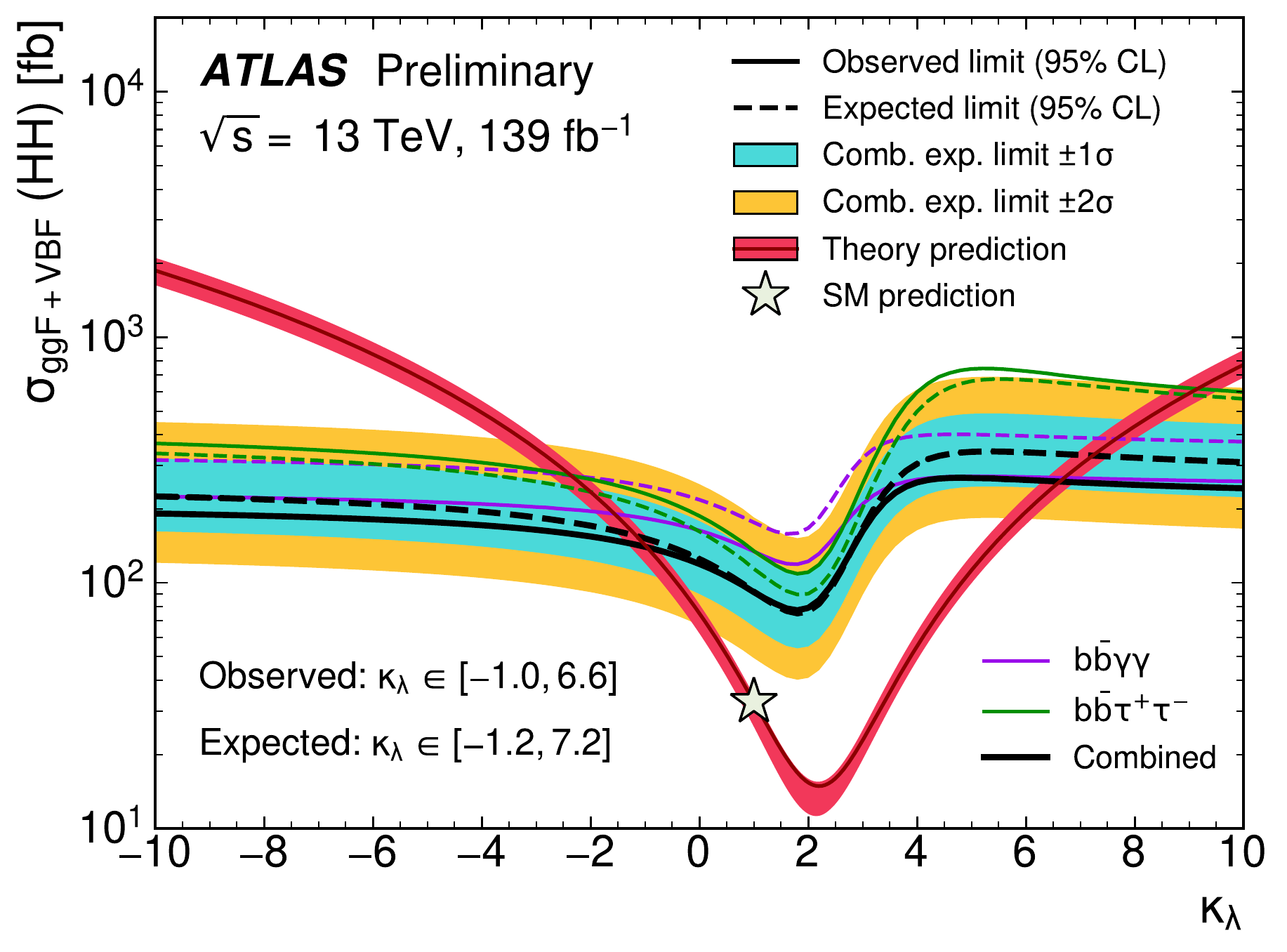}}
\end{minipage}
\hfill
\begin{minipage}[t]{0.485\textwidth}
\centerline{\includegraphics[width=0.95\textwidth]{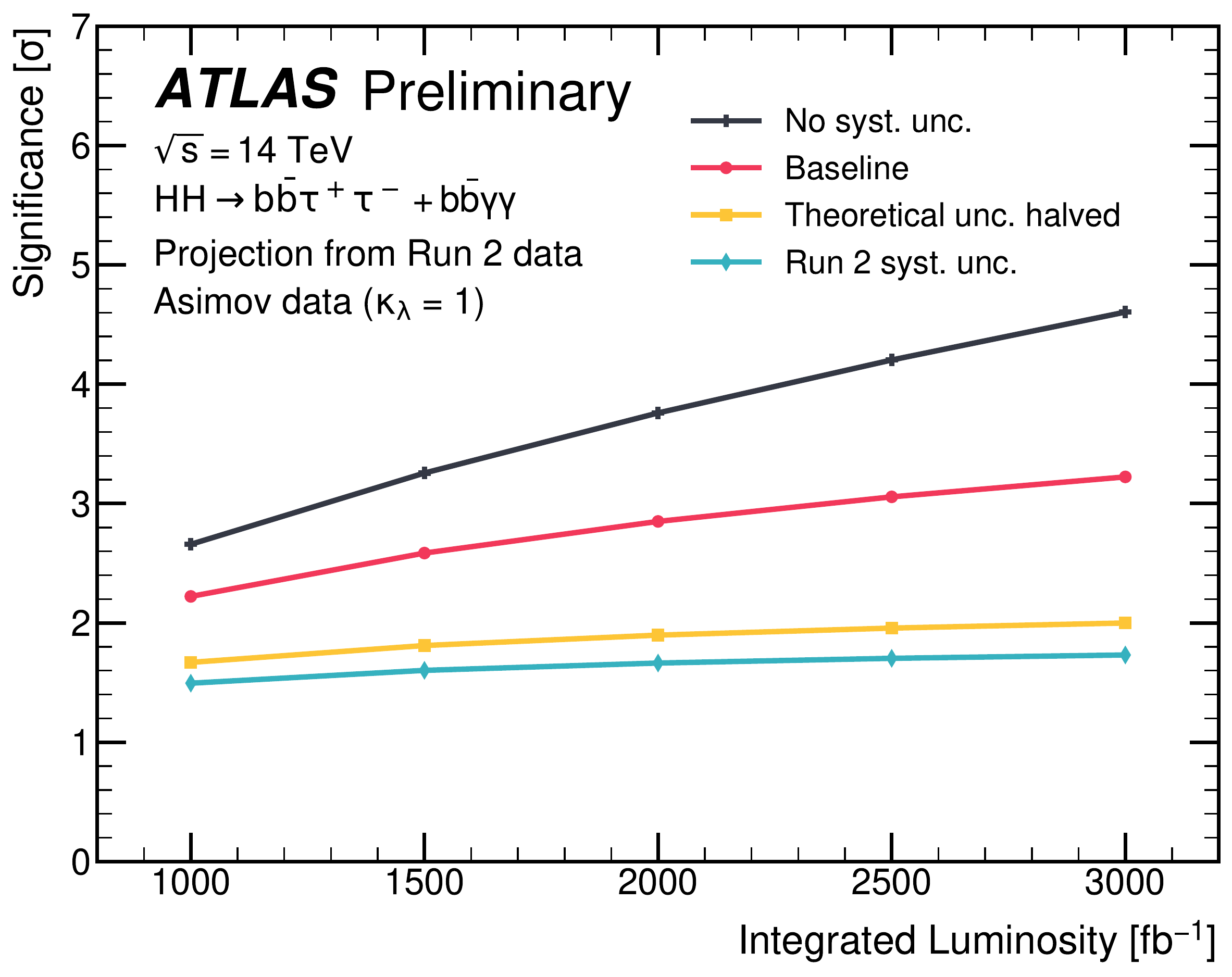}}
\end{minipage}
\vspace{-20pt}
\caption[]{Left: bounds on the di-Higgs production cross-section and trilinear
  coupling modifier ($\kappa_\lambda$) with the Run2 data~\cite{ATLAShh}. Right:
  expected significance of
  di-Higgs searches at the HL-LHC~\cite{ATLAShhproj}.}
\label{fig:HH}
\end{figure}

\noindent The measured cross-sections can be interpreted in terms of bounds on
$\kappa_\lambda$. The observed 95\% CL bound is: $\kappa_\lambda
\subset$ [-1.0, 6.6]~\cite{ATLAShh}, as shown in Figure~\ref{fig:HH} (left).

\noindent Searches with Run2 data 
are limited by statistical uncertainty. To gauge the potential for di-Higgs
observation, it is therefore important to
extrapolate  the sensitivity to the future high-luminosity LHC (HL-LHC), as
shown in Figure~\ref{fig:HH} (right). If no systematic uncertainty is
considered, the \hihi{} signal strength would be measured with 23\% statistical uncertainty
with the HL-LHC data-set of 3000~\ifb. Accounting for systematic uncertainty,
with a baseline estimate of how this uncertainty will evolve, the signal
strength is projected to be measured with $^{+34\%}_{-31\%}$ uncertainty,
constraining trilinear coupling modifier to $\kappa_\lambda
\subset$ [0.5,\,1.6] at 1\,$\sigma$ CL~\cite{ATLAShhproj}. 

\section{New Results for MoriondQCD}
\label{sec:new}

ATLAS results made available for the first time for the MoriondQCD confence
include: a new probe of the charge conjugation and parity (CP) of the top-Higgs coupling~\cite{ATLAStthcp}, and a measurement of the fiducial and differential cross-section of \VH{} production~\cite{ATLASvbbfid}.

The SM predicts the top-quark Yukawa coupling ($y_t$) to be CP-even, however a
presence of a CP-odd admixture has not yet been excluded. With a CP mixing angle
$\alpha$ the contribution to the Lagrangian is parameterised as:
\begin{equation}
\mathcal{L} =-y_t\overline{\Psi}_t\kappa'_t(\cos(\alpha)+i\sin(\alpha)\gamma^5)
         {\Psi}_t\Phi.
\label{eq:top_cp}
\end{equation}
The new analysis extracts $\alpha$ and the coupling strength modifier
$\kappa'_t$ and targets the final states with
top quarks and $\higgs\rightarrow\bbbar$ decays. The main challenge is
constraining non-resonant $\ttbar+\bbbar$ background, which 
dominates the measurement uncertainty. 
The angle $\alpha$ is extracted from fits to CP-sensitive observables, such as:
      \begin{equation}
      b_4=\frac{p^{z}_t
        p^{z}_{\overline{t}}}{|\vec{p}_t||\vec{p}_{\overline{t}}|},
      \label{eq:top_cp_b4}
    \end{equation}
where $p^{z}$ and $\vec{p}$ are the longitudinal momentum and the momentum
three-vector of the top and anti-top quark candidates. Figure \ref{fig:ttH} (left) shows the
distribution of the variable $b_4$ for CP-even ($\alpha=0^\circ$) and pure CP-odd
($\alpha=90^\circ$) hypotheses. In the CP-odd case, the top and anti-top are
more likely to fly in the opposite directions, therefore $b_4$ 
 is more likely to be negative. The data disfavours the pure CP-odd coupling at $1.2
\,\sigma$ significance. The best-fit value, shown in Figure \ref{fig:ttH} (right), is: $\alpha = {11^\circ} ^{+56^\circ}_{-77^\circ}.$ This is consistent with the
previously published ATLAS measurement of the CP-properties of the top-Higgs coupling in
$\higgs\rightarrow\gamma\gamma$ decays, which sets a constraint: $|\alpha| <
43^\circ $ at 95\% CL and excludes pure CP-odd coupling at $3.9\,\sigma$
significance~\cite{ATLAStthcpyy}.
\begin{figure}[!ht]
\begin{minipage}{0.5\textwidth}
\centerline{\includegraphics[width=0.92\textwidth]{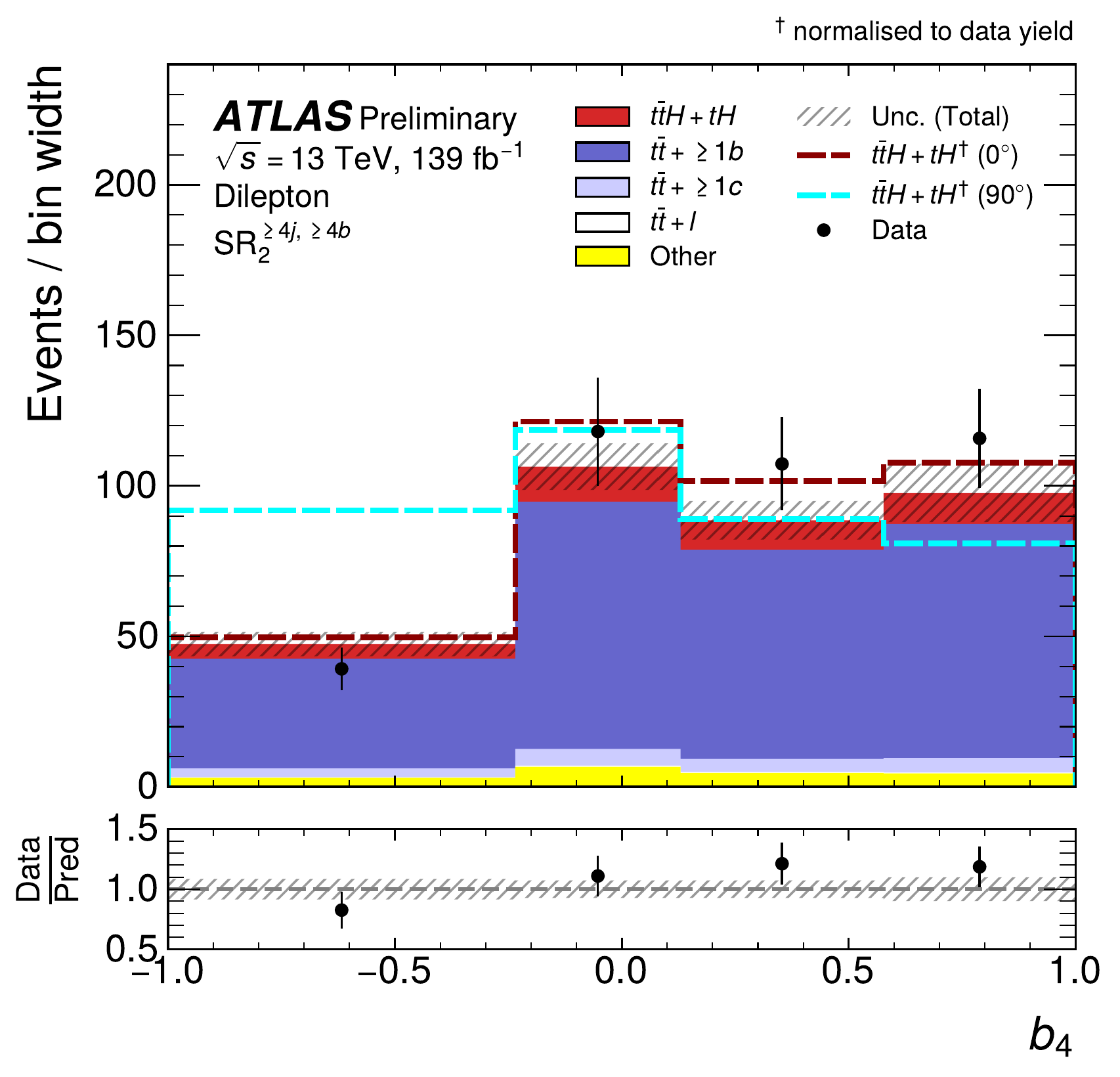}}
\end{minipage}
\hfill
\begin{minipage}{0.47\textwidth}
\vspace{-10pt}
\centerline{\includegraphics[width=0.92\textwidth]{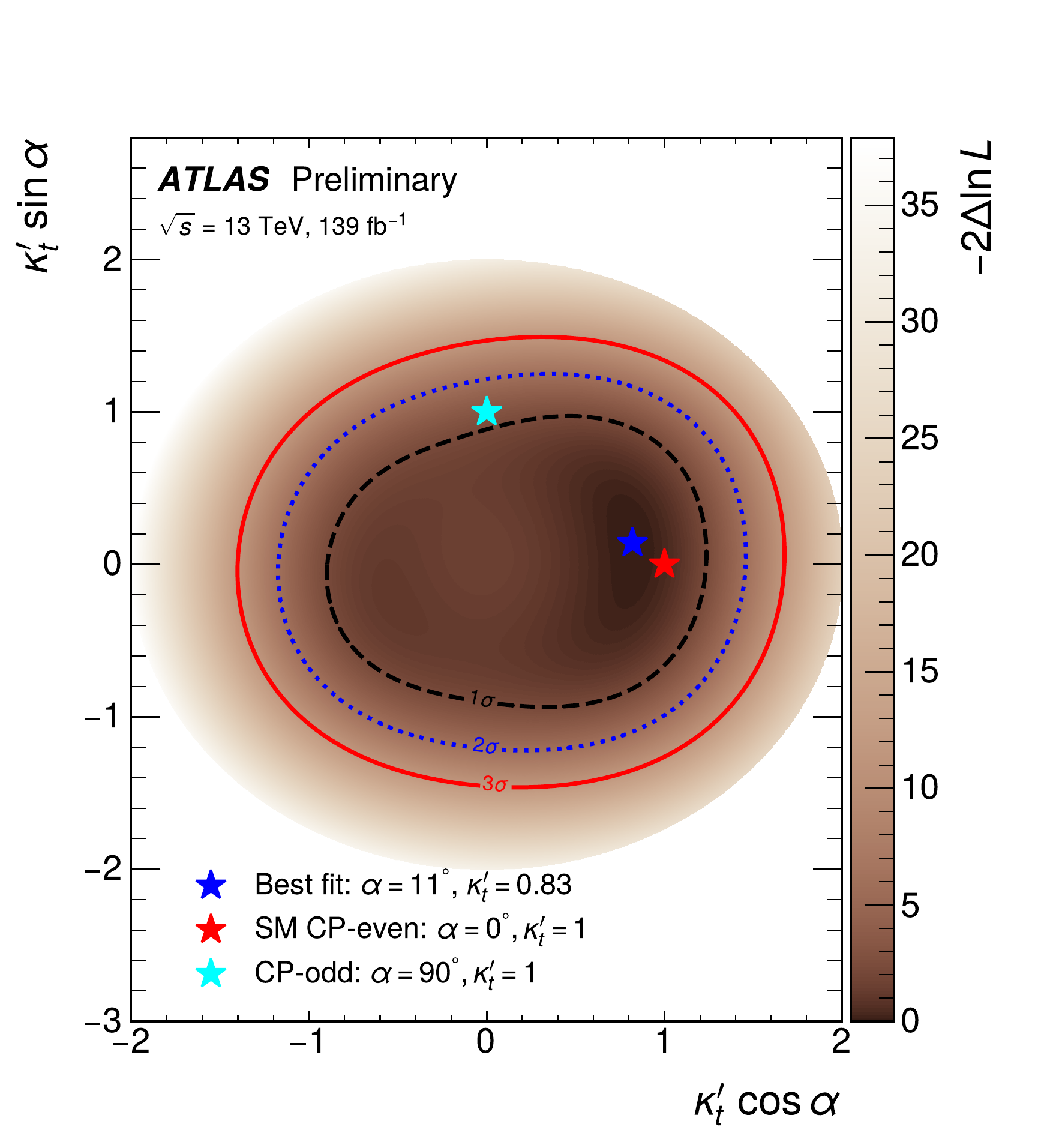}}
\end{minipage}
\vspace{-10pt}
\caption[]{Left: fit to observable $b4$, used to constrain the CP
  properties of the top-Higgs coupling. Right: the observed constraints on the coupling
  modifier ($\kappa'_t$) and the CP mixing angle ($\alpha$)~\cite{ATLAStthcp}.}
\label{fig:ttH}
\end{figure}

\noindent The new fiducial and differential cross-section measurement of
the \VH{} production targets $\higgs\rightarrow\bbbar,Z\rightarrow\nu\nu$ decays.
The idea behind the {\sl fiducial} measurements is to
use particle-level selection criteria close to the detector acceptance. 
Such measurements require no extrapolation outside the measured
phase-space, which minimises model dependence.
The new measurement extracts the signal strengths 
from a fit to the invariant mass $m_{\bbbar}$, shown in Figure \ref{fig:vhfid}
(left). The neutrinos are reconstructed as missing transverse 
energy \etmiss, and the signal strengths are extracted in two bins: 
$150\GeV {<} \etmiss {<} 250 \GeV$ and  $\etmiss {>} 250 \GeV$, as shown in
Figure~\ref{fig:vhfid} (right).
      
\begin{figure}[!ht]
 \vspace{-10pt}
\begin{minipage}{0.4\textwidth}
\centerline{\includegraphics[width=1.00\textwidth]{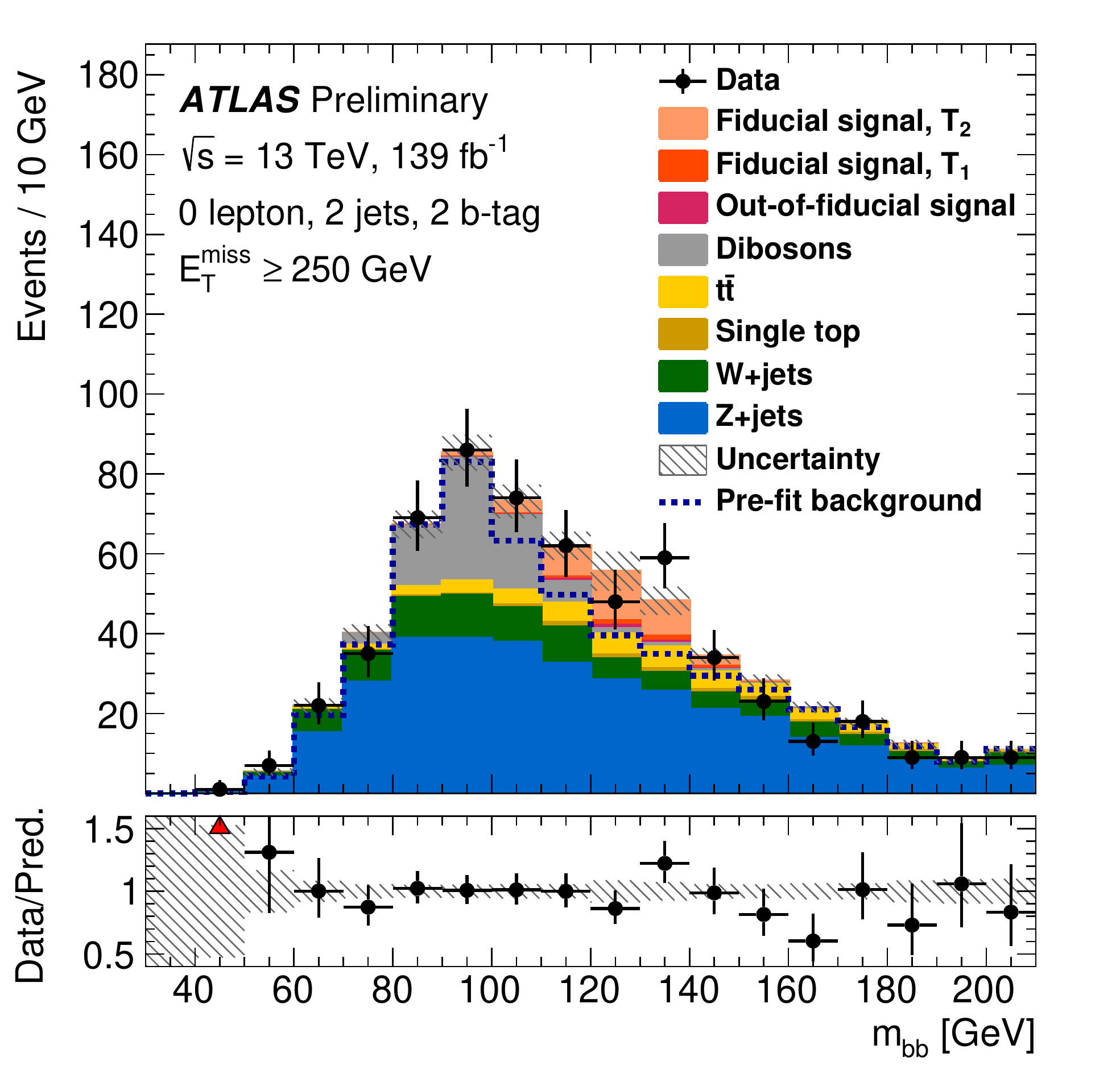}}
\end{minipage}
\hfill
\begin{minipage}{0.6\textwidth}
\centerline{\includegraphics[width=1.00\textwidth]{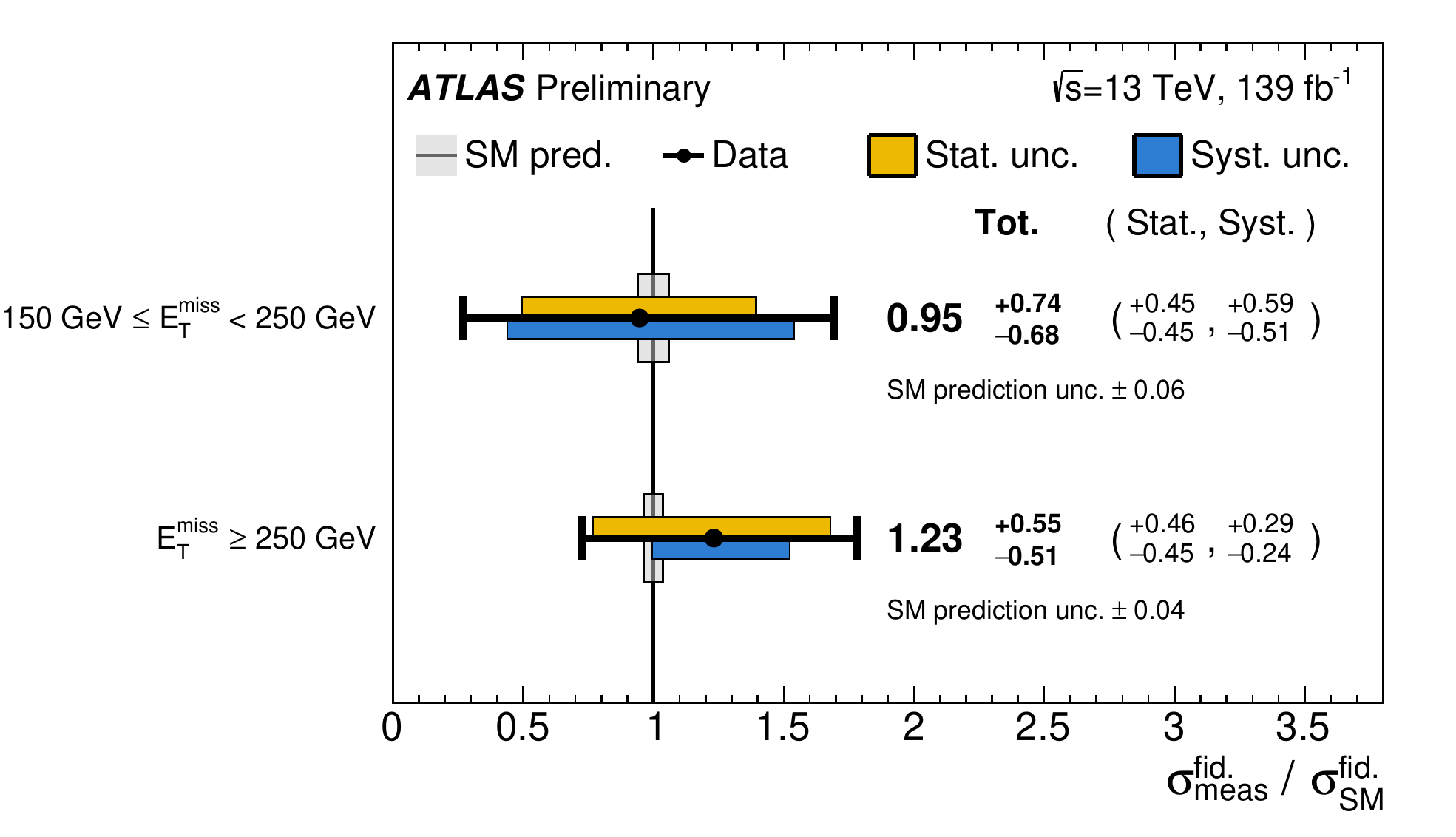}}
\end{minipage}
\vspace{-10pt}
\caption[]{Left: invariant mass spectrum fit, used to measure
  $\VH,{\higgs}{\rightarrow}{\bbbar}$ fiducial cross-sections.
  The T1 and T2 denote two \etmiss{} bins: $150\GeV {<} \etmiss {<} 250 \GeV$ and 
  $\etmiss {>} 250 \GeV$. Right: the measured signal strengths~\cite{ATLASvbbfid}.}
\label{fig:vhfid}
\end{figure}

\noindent The new fiducial measurement is complementary to previously published ATLAS
$\VH,\higgs\rightarrow\bbbar$ results, which measure the cross-section
extrapolated to simplified volumes in so-called {\sl Simplified Template
  Cross-Section} (STXS) framework~\cite{ATLASvbb}. These measure the $WH$ and $ZH$ production
cross-sections with 4.0\,$\sigma$ and 5.3\,$\sigma$ significance respectively. 

\section{Summary}
Within a decade of the Higgs boson discovery, ATLAS has observed all
main production modes and measured the inclusive production cross-section with
7\% uncertainty.  The quest for new experimental probes of the Higgs mechanism
continues and measurements with the LHC Run2 data have advanced our knowledge.
New break-throughs in answering the major open
questions are: (1) a measurement confirming that the
charm-quark Yukawa coupling is smaller than the $b$-quark coupling; (2) enhanced
sensitivity of di-Higgs production searches, with an extrapolated di-Higgs
cross-section measurement sensitivity of $^{+34\%}_{-31\%}$ at the HL-LHC. 

\end{document}
